\documentclass[aps,prd,preprint,superscriptaddress]{revtex4}
\usepackage{graphicx}
\usepackage{amsmath}
\usepackage{bm}% bold math
\usepackage{color}% bold math
\usepackage{amssymb,enumerate}

\def\=:{=\hspace{-.7em}\raisebox{1.1ex}{.}\hspace{.1em}\raisebox{-0.2ex}{.} }

\newcommand {\del}{\partial}

\begin{document}

% Use the \preprint command to place your local institutional report
% number in the upper righthand corner of the title page in preprint mode.
% Multiple \preprint commands are allowed.
% Use the 'preprintnumbers' class option to override journal defaults
% to display numbers if necessary
%\preprint{}

%Title of paper
\title{
${\mathbb Z}_n$ modified XY and Goldstone models\\
and vortex confinement transition
}

% repeat the \author .. \affiliation  etc. as needed
% \email, \thanks, \homepage, \altaffiliation all apply to the current
% author. Explanatory text should go in the []'s, actual e-mail
% address or url should go in the {}'s for \email and \homepage.
% Please use the appropriate macro foreach each type of information

% \affiliation command applies to all authors since the last
% \affiliation command. The \affiliation command should follow the
% other information
% \affiliation can be followed by \email, \homepage, \thanks as well.

\author{Michikazu Kobayashi}
\email{michikaz@scphys.kyoto-u.ac.jp}
\affiliation{Department of Physics, Kyoto University, Oiwake-cho, Kitashirakawa, Sakyo-ku, Kyoto 606-8502, Japan}

\author{Muneto Nitta}
\email{nitta(at)phys-h.keio.ac.jp}
\affiliation{
Department of Physics, and Research and Education Center for Natural 
Sciences, Keio University, Hiyoshi 4-1-1, Yokohama, Kanagawa 223-8521, Japan\\
}
%Collaboration name if desired (requires use of superscriptaddress
%option in \documentclass). \noaffiliation is required (may also be
%used with the \author command).
%\collaboration can be followed by \email, \homepage, \thanks as well.
%\collaboration{}
%\noaffiliation

%Collaboration name if desired (requires use of superscriptaddress
%option in \documentclass). \noaffiliation is required (may also be
%used with the \author command).
%\collaboration can be followed by \email, \homepage, \thanks as well.
%\collaboration{}
%\noaffiliation

\date{\today}
\begin{abstract}
The modified XY model is a modification of the XY model 
by addition of a half-periodic term.
The modified Goldstone model is a regular and continuum version of the modified XY model. 
The former admits a vortex molecule, that is, two half-quantized vortices connected by a domain wall, as a regular topological soliton solution to the equation of motion while the latter admits it as a singular configuration.
Here we define the ${\mathbb Z}_n$ modified XY and Goldstone models 
as the $n=2$ case to be the modified XY and Goldstone models, 
respectively. 
We exhaust all stable and metastalble vortex solutions for $n=2,3$ 
and find a vortex confinement transition from an integer vortex to a vortex molecule 
of $n$ $1/n$-quantized vortices, 
depending on the ratio between the term of the XY model 
and the modified term.
We find for the case of $n=3$, a rod-shaped molecule is the most stable while a Y-shaped molecule is 
metastable. 
We also construct some solutions for the case of $n=4$.
The vortex confinement transition can be understood in terms of 
the  ${\mathbb C}/{\mathbb Z}_n$ orbifold geometry.

\end{abstract}
% insert suggested PACS numbers in braces on next line
% insert suggested keywords - APS authors don't need to do this
%\keywords{}

%\maketitle must follow title, authors, abstract, \pacs, and \keywords
\maketitle

\section{Introduction}
The XY model is a lattice model 
describing a lot of physical systems such as 
superconductors and superfluids.
Its Hamiltonian is given by  
$
{\cal H}_{\rm XY} =
-J \sum_{\langle i,j \rangle} \cos (\vartheta_i - \vartheta_j),
$ 
where $i,j$ label the lattice cites 
and ${\langle i,j \rangle} $ implies a pair of nearest neighbours.
In the continuum limit, it becomes just a free U(1) scalar field theory 
or nonlinear O(2) model. 
One of the most nontrivial features of this model is to exhibit 
a topological phase transition called 
the Berezinskii-Kosterlitz-Thouless (BKT) transition 
\cite{berezinskii71,berezinskii72,kosterlitz71,kosterlitz72} 
in 2+1 dimensions, 
which separates bound vortices at low temperature 
and liberated pairs of vortex and anti-vortex 
at high temperature.
The BKT transition yields quasi-long-range order with algebraically decaying correlations,
although long-range order with continuous symmetry is forbidden by the Coleman-Mermin-Wagner (CMW) theorem \cite{Coleman:1973ci,mermin66,Hohenberg:1967zz}.
The BKT transition has been confirmed experimentally in various condensed matter systems 
such as 
$^4$He films \cite{Bishop}, thin superconductors \cite{Gubser,Hebard,Voss,Wolf,Epstein}, Josephson-junction arrays \cite{Resnick,Voss2}, colloidal crystals \cite{Halperin,Young,Zahn,Nakamura}, and ultracold atomic Bose gases \cite{Hadzibabic}. 
One of drawbacks of the XY model may be the fact that 
vortices are singular configurations but not solutions to 
the equation of motion. 
To overcome this problem, 
one can introduce a Higgs (amplitude) degree of freedom together with
 a potential term along the Higgs direction, 
 then the model becomes the Goldstone or linear O(2) model, 
%Vortices in the XY or nonlinear O(2) model is singular configuration, 
%but the Goldstone model or linear O(2) model 
allowing 
vortices as regular solutions to the equation of motion.
The vortex core singularity is resolved by the Higgs field 
while the large distance behavior can be capture by the XY model.

The modified XY model is a modification of the XY model 
by addition of a half-periodic term 
\cite{domany84,korshunov85,lee85,carpenter89}:
\begin{align}
{\cal H}_{\rm mXY} = -J \sum_{\langle i,j \rangle} \cos (\vartheta_i - \vartheta_j) - J^\prime \sum_{\langle i,j \rangle} \cos [2(\vartheta_i-\vartheta_j)],
\label{Eq:modXY}
\end{align} 
where the second term is the half-periodic term. 
This model admits a vortex molecule, that is, two half-quantized vortices connected by a domain wall, as a singular configuration,
and its existence is crucial in the phase diagram, 
as is so for the XY model.
When the coupling $J'$ of the modified term is large enough compared with the coupling $J$, 
there exists an Ising type phase transition \cite{korshunov85,lee85,carpenter89}  
as a consequence of the presence of domain walls.
The modified model in Eq.(\ref{Eq:modXY}) and its various modifications \cite{dian11,shi11,bonnes12,huebscher13,serna17,nui18,canova16,zukovic17,zukovic18} 
are of great importance and interest, because of applicability to various systems such as superfluidity in atomic Bose gases \cite{radzihovsky08}, arrays of unconventional Josephson junctions \cite{korshunov10}, or high temperature superconductivity \cite{komendova10}.  
The modified Goldstone or modified linear O(2) model is a regular (complemented by the Higgs mode) and continuum version of the modified XY model
\cite{Kobayashi:2019sus}:
\begin{align}
\label{Eq:Ham}
{\cal H}_{\rm mGoldstone}  &= \int d^d x \left[ a |\nabla \phi|^2 + b |\nabla \phi^2|^2 + \frac{\lambda}{2} \left( |\phi|^2 - v^ 2\right)^2 \right],
\end{align}
where $\phi = \exp r (i \vartheta)$ is a complex scalar field ($r$ is the Higgs field), and $\lambda$, $a$ and $b$ are positive coupling constants 
determined from the lattice model. 
%The continuum version of the standard XY model refers to $b=0$ and in the modified XY model we have $b> 0$. 
This model admits a vortex molecule of half-quantized vortices connected by a domain wall 
as a regular topological soliton solution to the equation of motion 
when $b$ is large enough \cite{Kobayashi:2019sus},
while for small $b$ the molecule collapses to an integer vortex. 
The phase diagram is quite rich and there is a two-step phase transition of BKT type and 
of Ising type \cite{Kobayashi:2019sus}.

In this paper, as a generalization of the modified XY and Goldstone models, 
we define the ${\mathbb Z}_n$ modified XY and Goldstone models:
\begin{align}
{\cal H}_{{\mathbb Z}_n{\rm mXY}} = -J \sum_{\langle i,j \rangle} \cos (\vartheta_i - \vartheta_j) - J^\prime \sum_{\langle i,j \rangle} \cos [n(\vartheta_i-\vartheta_j)],
\label{Eq:modZnXY}
\end{align} 
and
\begin{align}
\label{Eq:modZnHam}
{\cal H}_{{\mathbb Z}_n{\rm mGoldstone}}  &=  \int d^d x \left[ a |\nabla \phi|^2 + b |\nabla \phi^n|^2 + \frac{\lambda}{2} \left( |\phi|^2 - v^2 \right)^2 \right],
\end{align}
respectively. The case of $n=2$ corresponds to the usual modified XY and Goldstone models.
We study vortex solutions in this model
with particular attension to the cases of $n=2,3$. 
We exhaust stable and metastable vortex solutions for  these cases, 
and find a vortex confinement transition from an integer vortex to a vortex molecule, 
depending on the ratio between $a$ and $b$. 
We find for the case of $n=3$ that a rod-shaped molecule is the most stable while a Y-shaped molecule is 
metastable. 
We also give some examples of (meta)stable vortices in the case of $n=4$.
This transition can be understood in terms of the 
${\mathbb C}/{\mathbb Z}_n$ orbifold geometry;
the model can be written in the form of a nonlinear sigma model with the target space 
${\mathbb C}/{\mathbb Z}_n$ with a possible orbifold singularity resolved. 
If vacua are at large distance from the origin in the target space, 
a vortex becomes a molecule of $1/n$ fractional vortices, 
while if  the vacua are close to the origin the vortex becomes an integer vortex.

This paper is organized as follows. 
In Sec.~\ref{sec:model}, 
we introduce our model and discuss geometry.
In Sec.~\ref{sec:vortices}, we construct vortex solutions.
Section \ref{sec:summary} is devoted to a summary 
and discussion. 

\newpage

%%%%%%%%%%%%%%%%%%%%%%%%%%
\section{The model and geometry \label{sec:model}}
In this section, we formulate our model and discuss geometric properties.
The Lagrangian of the ${\mathbb Z}_n$ modified Goldstone model is given by
\begin{align}
\begin{split}
  {\cal L} &= a \del_\mu \phi^\ast \del^\mu \phi 
                     + \frac{b}{n} \del_\mu \phi^{*n} \del^\mu \phi^n  
                     - \frac{\lambda}{2} (|\phi|^2 -v^2)^2 \\
            &= (a + b n |\phi^{n-1}|^2)  \del_\mu \phi^\ast \del^\mu \phi  
                - \frac{\lambda}{2} (|\phi|^2 -v^2)^2.
\end{split}
\label{eq:Lagrangian}
\end{align}
The vacua are  $S^1$ defined by $|\phi|^2 = v^2$.
This model is just a nonlinear sigma model with the target space metric
\begin{align} 
g(\phi,\phi^\ast)=  a + b n |\phi^{n-1}|^2.  \label{eq:metric}
\end{align}
In the limit of  $\lambda \to \infty$, the model reduces to 
an $O(2)$ nonlinear sigma model (or the XY model) with the Lagrangian
${\cal L} = (a + b n v^{2n-2})  \del_\mu \phi^* \del^\mu \phi$ with a constraint 
$|\phi|^2 =v^2$.
It is sometimes useful to rewrite the Lagrangian by a new field $\Phi = \phi^n$ as
\begin{align}
\begin{split}
  {\cal L} &= a \del_\mu \Phi^{\ast 1/n} \del^\mu \Phi^{1/n} 
                     + \frac{b}{n} \del_\mu \Phi^* \del^\mu \Phi  
                     - \frac{\lambda}{2} (|\Phi^{1/n}|^2 -v^2)^2 \\
            &= \frac{1}{n} \left( \frac{a}{n} |\Phi^{- \frac{n-1}{n}}|^2 + b \right)  \del_\mu \Phi^\ast \del^\mu \Phi  
                     - \frac{\lambda}{2} (|\Phi^{1/n}|^2 -v^2)^2 .
\end{split}
\label{eq:Lagrangian2}
\end{align}

Let us discuss the asymptotic behaviour of the target space geometry.
Writing $\phi = r e^{i \theta}$ or $\Phi = R e^{i \Theta}$ ($R=r^n$ and $\Theta = n \theta$),
the geometry behaves two different ways separated by the critical radius 
$r=r_c$ defined by
\begin{align}
 r_c = \left(\frac{a}{bn}\right)^\frac{1}{2n-2}, \quad
  R_c = \left(\frac{a}{b n}\right)^\frac{n}{2n-2}. \
\end{align}
Then, we can see that the metric behaves differently at large and short distances as follows:
\begin{itemize}
\item
For the large distance, $r \gg r_c$  $(R \gg R_c)$, the first term in the metric |
in Eq.~\eqref{eq:metric} is negligible and the Lagrangian reduces to
\begin{align} 
\begin{split}
{\cal L}_{\rm large} 
  &= \frac{b}{n}  \del_\mu \Phi^\ast \del^\mu \Phi  - \frac{\lambda}{2} (|\Phi^{1/n}|^2 - v^2)^2 \\
  &= \frac{b}{n} \del_\mu \phi^{\ast n} \del^\mu \phi^n  
                     - \frac{\lambda}{2} (|\phi|^2 -v^2)^2 .
\end{split}
\label{eq:Lag-large}
\end{align}
One observes that $\Phi$ is a good coordinate rather than $\phi$.
The kinetic term of the Lagrangian in Eq.~\eqref{eq:Lag-large}  is just a free scalar field in terms of $\Phi$, but the target space is 
rather an orbifold:
\begin{align}
 {\cal M} \simeq {\mathbb C}/{\mathbb Z}_n.
\end{align}
This is because 
 all $\phi \omega^a$ with $a=0,1,2,\cdots n-1$ 
yield the same $\Phi$,
where $\omega^n=1, \omega = \exp (2\pi i/n)$.
%The vacua are
%\begin{align}
%\Phi^{1/n} = v e^{i \alpha}, 
%\end{align}
This metric has an orbifold singularity in the origin, 
but it is not the case for the whole metric.
\item
In fact, the short distance behaviour ($r \ll r_c$  $(R \ll R_c)$) is dominated by the first term in the metric, and the Lagrangian reduces to
\begin{align}
\begin{split}
  {\cal L}_{\rm short} 
  &=  a \del_\mu \phi^\ast \del^\mu \phi 
                     - \frac{\lambda}{2} (|\phi|^2 -v^2)^2 \\
  &= a \del_\mu \Phi^{\ast 1/n} \del^\mu \Phi^{1/n}
                     - \frac{\lambda}{2} (|\Phi^{1/n}|^2 -v^2)^2.
\end{split}
\end{align}
The Lagrangian is nothing but the usual Goldstone model in terms of $\phi$. 
In this case, $\phi$ is a good coordinate in which 
the metric is smooth at the origin $\phi=0$.
Therefore, we have seen that a possible singularity 
in the orbifold 
${\mathbb C}/{\mathbb Z}_n$ is resolved in the full metric, and the whole target space is smooth.

\end{itemize}

\section{Vortices} \label{sec:vortices}
Comparing the vacua $r=v$ and the critical radius $r=r_c$ around which the geometry behaves differently, we find two different scheme of 
the structure of vacua and consequently that of vortices. 
When the vacua exist inside the critical radius $r=r_c$, that is 
$v \ll r_c$, we do not need the outside geometry,
in which case the Lagrangian reduces to the usual Goldstone model of $\phi$ admitting 
the $S^1$ vacua and integer global vortices.
A single vortex configuration is of the form of $\phi = f(\rho) \exp (i\varphi)$ 
with the polar coordinates $(\rho,\varphi)$.
  
On the other hand, when $r_c \ll v$, 
the Lagrangian is well described by $\Phi$ in 
Eq.~\eqref{eq:Lagrangian2}, 
which is asymptotically reducing the Lagrangian  Eq.~(\ref{eq:Lag-large})
 of 
a Goldstone model in terms of $\Phi$. 
The vacua are $\Phi^{1/n} = v e^{i \alpha}$. 
(There remain ${\mathbb Z}_n$? 
since $\phi \sim \phi \omega^a$ yield the same $\Phi$ with
$\omega^n=1, \omega = \exp (2\pi i/n)$.
If $a=0$,  $\Phi$ is always a good coordinate and 
the model admits $1/n$ quantized (fractional) global vortices 
$\Phi = g^n(\rho) \exp (i \varphi)$, 
[$\phi = g(\rho) \exp (i \varphi/n)$].
However, if $a\neq 0$, and these fractional vortices cannot exist alone, 
since only $\phi$ is a good coordinate in the vicinity of the origin of the target space. 
Instead, $n$ of them must be confined to one integer vortex.

In summary, we have the following two cases:
%For general case with nonzero $a$ and $b$, the integer vortex may be decomposed into $n$ fractional vortices. 
\begin{itemize}
\setlength{\parskip}{0pt}
\setlength{\itemsep}{0pt}
\item
$r_c \gg v$, $R_c \gg v^n$: integer vortex scheme.   
\item
$r_c \ll v$, $R_c \ll v^n$:     fractional vortex scheme.            
\end{itemize}

We have numerically obtained the stationary solution with one integer vortex in 2-dimensional space by minimizing the energy
\begin{align}
\mathcal{E} = \int_\Omega dx^2\: \left\{ a |\nabla \phi|^2 
                                 + \frac{b}{n} |\nabla \phi^n|^2  
                     + \frac{\lambda}{2} (|\phi|^2 - v^2)^2 \right\}.
\label{eq:vortex-energy}
\end{align}
The solution can be calculated by finding the solution
\begin{align}
\begin{split}
0 = \frac{\delta \mathcal{E}}{\delta \phi^\ast}
  = - a \nabla^2 \phi - b (\nabla^2 \phi^n) \phi^{\ast n-1}
                     + \lambda (|\phi|^2 - v^2) \phi,
\label{eq:modified-GP}
\end{split}
\end{align}
under the boundary condition $\phi = v e^{i \varphi}$ (on $\del\Omega$).
As a numerical parameters, we have chosen $\lambda = v = 1$.
$a$ and $b$ are parametrized by $\theta$ as $a = \cos\theta$ and $b = \sin\theta$.
\begin{figure}[htb]
\centering
\includegraphics[width=0.5\linewidth]{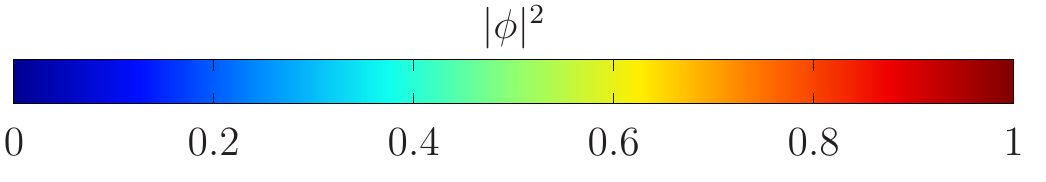} \\
\begin{minipage}{0.24\linewidth}
{\footnotesize (a) $\theta = 82^\circ$} \\
\includegraphics[width=0.99\linewidth]{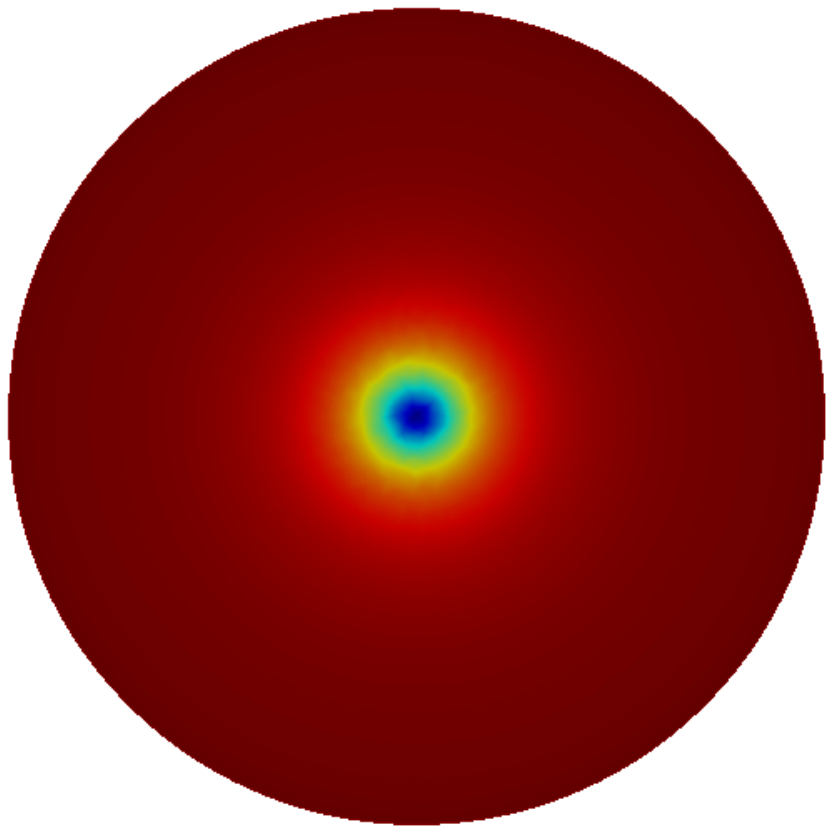}
\end{minipage}
\begin{minipage}{0.24\linewidth}
{\footnotesize (b) $\theta = 84^\circ$} \\
\includegraphics[width=0.99\linewidth]{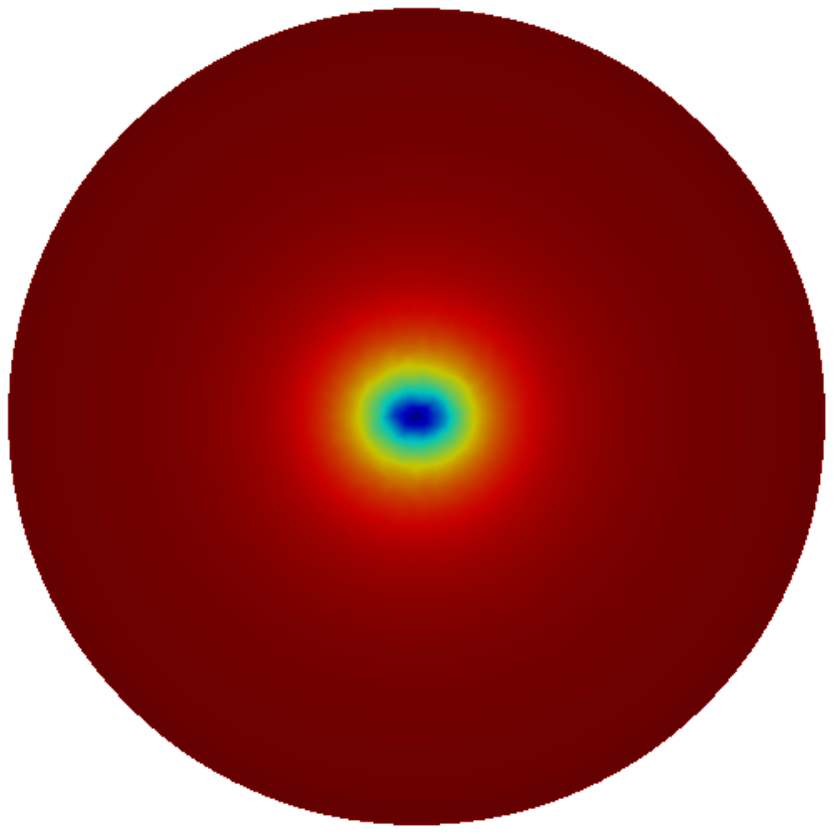}
\end{minipage}
\begin{minipage}{0.24\linewidth}
{\footnotesize (c) $\theta = 86^\circ$} \\
\includegraphics[width=0.99\linewidth]{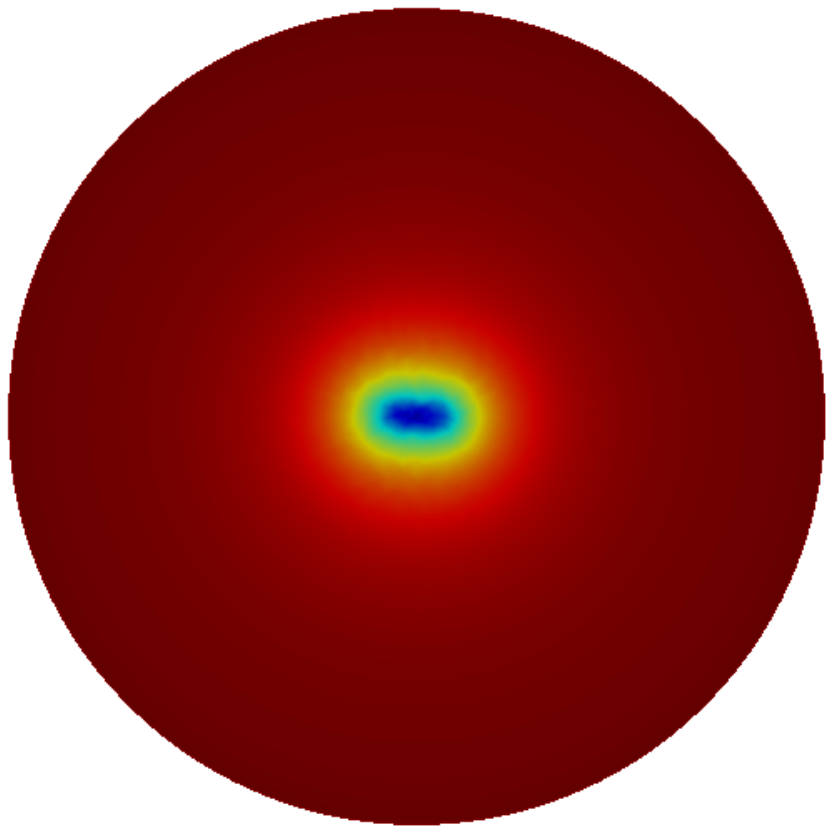}
\end{minipage}
\begin{minipage}{0.24\linewidth}
{\footnotesize (d) $\theta = 88^\circ$} \\
\includegraphics[width=0.99\linewidth]{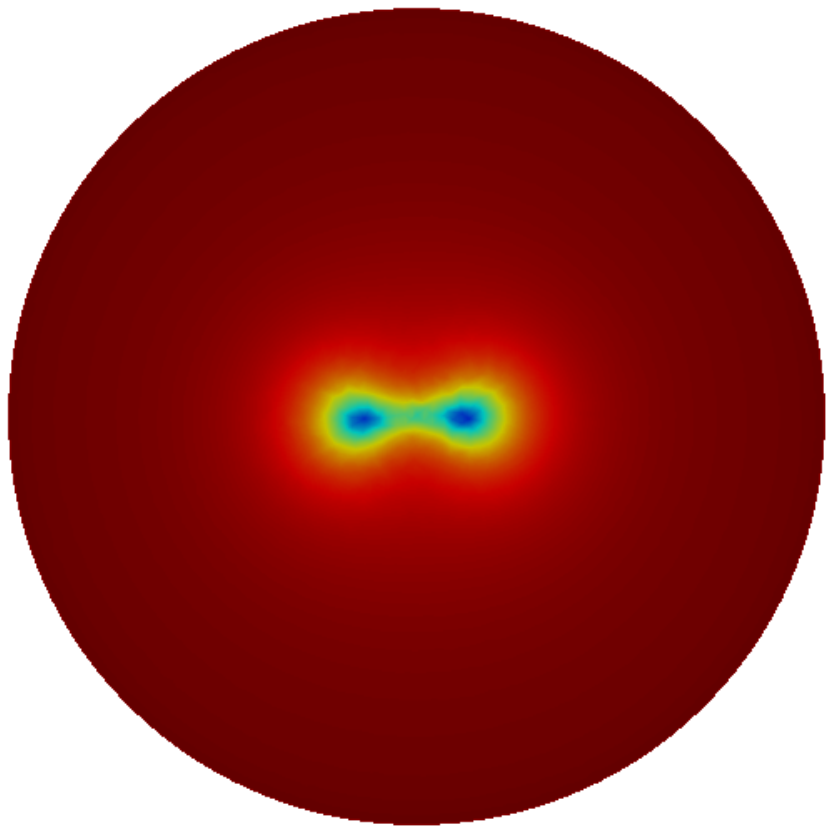}
\end{minipage}
\caption{
\label{fig:vortex-2}
Spatial configuration of $|\phi|^2$ for a integer vortex with $n = 2$.
The radius of the system is $20$.
}
\end{figure}
\begin{figure}[htb]
\centering
\includegraphics[width=0.5\linewidth]{colorbox.pdf} \\
\begin{minipage}{0.24\linewidth}
{\footnotesize (a) $\theta = 80^\circ$} \\
\includegraphics[width=0.99\linewidth]{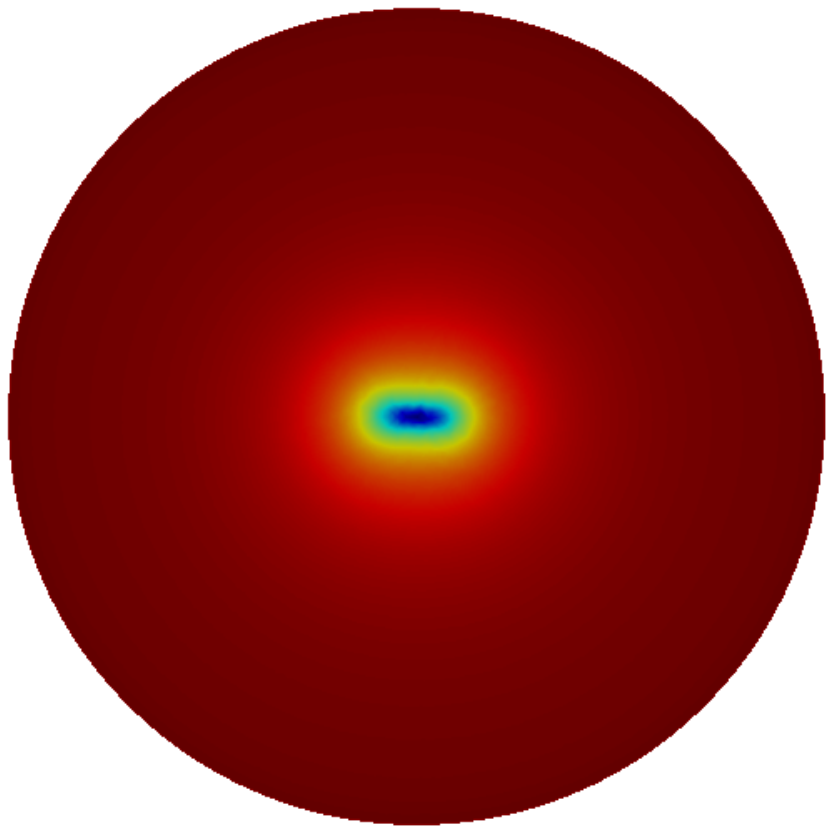}
\end{minipage}
\begin{minipage}{0.24\linewidth}
{\footnotesize (b) $\theta = 84^\circ$} \\
\includegraphics[width=0.99\linewidth]{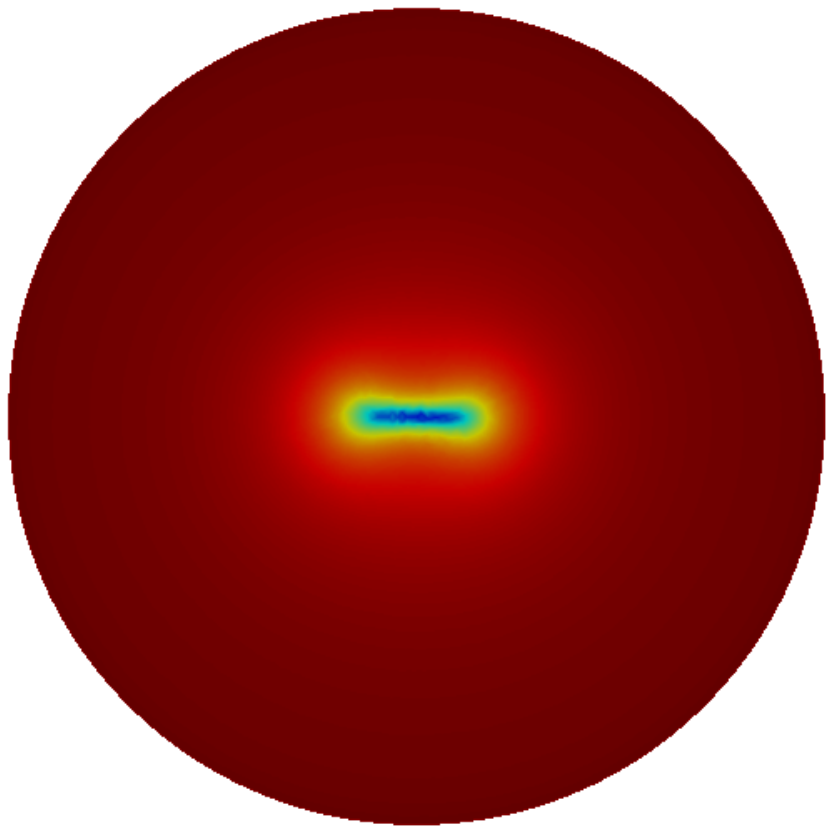}
\end{minipage}
\begin{minipage}{0.24\linewidth}
{\footnotesize (c) $\theta = 88^\circ$} \\
\includegraphics[width=0.99\linewidth]{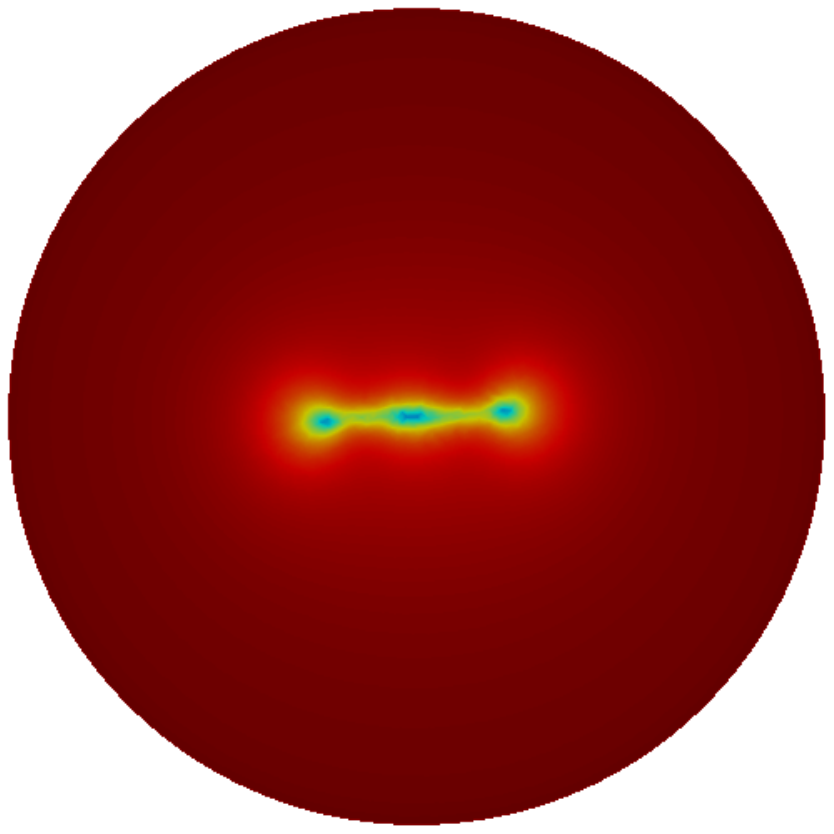}
\end{minipage}
\begin{minipage}{0.24\linewidth}
{\footnotesize (d) $\theta = 88^\circ$} \\
\includegraphics[width=0.99\linewidth]{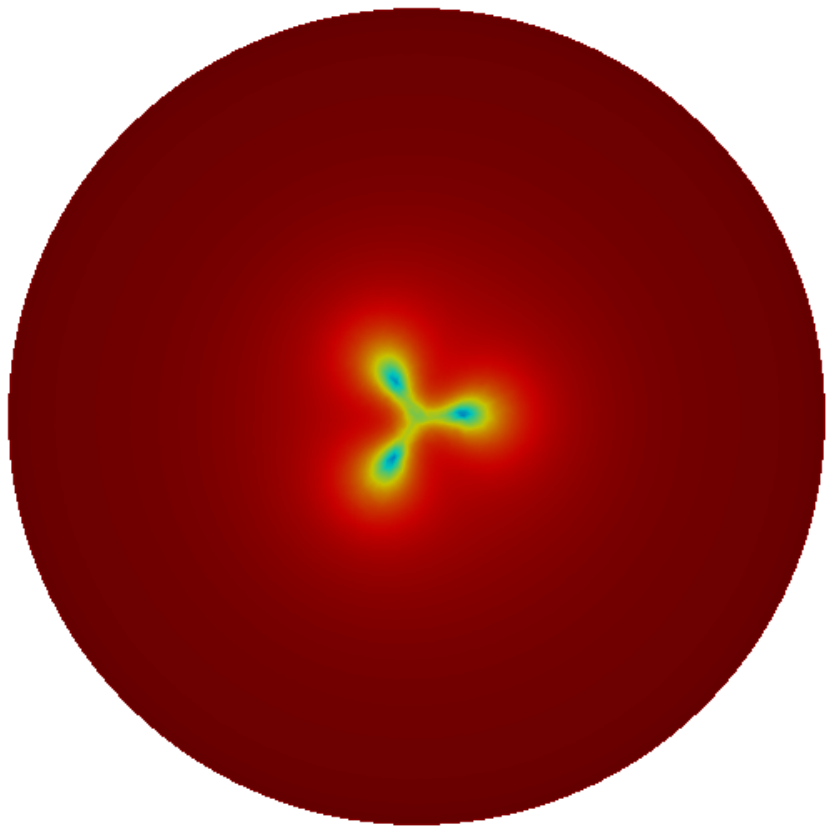}
\end{minipage}
\caption{
\label{fig:vortex-3}
Spatial configuration of $|\phi|^2$ for a integer vortex with $n = 3$.
The radius of the system is $30$.
While the solution in panel (c) is the stable ground state, the solution in panel (d) is the metastable state having the higher energy than that in panel (c). 
}
\end{figure}
Figures \ref{fig:vortex-2} and \ref{fig:vortex-3} show the spatial configuration of $|\phi|^2$ for a integer vortex with $n = 2$ and $n = 3$, respectively.
For small $\theta$, the vortex has the circular structure which is qualitatively the same as that for the usual Goldstone model with $\theta = 0$.
On the other hand, the circular integer vortex becomes energetically unstable and splits into $n$ fractional vortices connected with line defects.
We call this structure a vortex molecule.
For $n = 3$, furthermore, there are several metastable solutions for 
Eq. \eqref{eq:modified-GP} when $\theta$ is large.
The triangular-shaped molecule in Fig. \ref{fig:vortex-3} (d) is metastable and has the higher relative energy $\bar{\mathcal{E}} \equiv \mathcal{E} - \mathcal{E}_{\rm sym} \sim - 1.184$ than the rod-shaped molecule in Fig. \ref{fig:vortex-3} (c) having the lowest relative energy $\mathcal{E} - \mathcal{E}_{\rm sym} \sim -1.837$, where $\mathcal{E}_{\rm sym}$ denote the energy for the symmetric solution satisfying $\phi = g(\rho) e^{i \varphi}$.

\begin{figure}[htb]
\centering
\begin{minipage}{0.4\linewidth}
{\footnotesize (a) $n = 2$} \\
\includegraphics[width=0.99\linewidth]{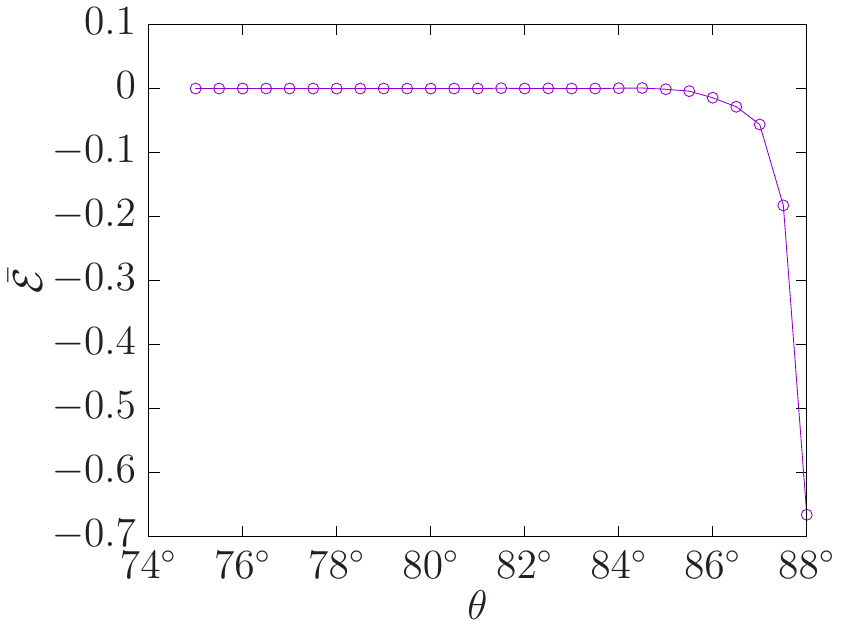}
\end{minipage}
\begin{minipage}{0.4\linewidth}
{\footnotesize (b) $n = 3$} \\
\includegraphics[width=0.99\linewidth]{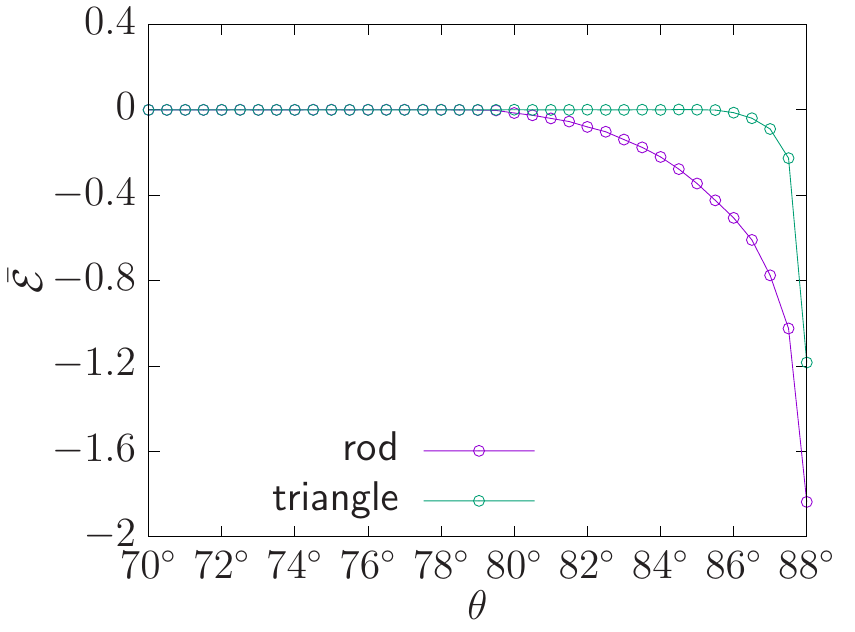}
\end{minipage}
\caption{
\label{fig:vorene}
Dependence of the relative energy $\bar{\mathcal{E}}$ on $\theta$.
}
\end{figure}
Figure \ref{fig:vorene} shows the dependence of the relative energy $\bar{\mathcal{E}}$ on $\theta$.
The zero relative energy $\bar{\mathcal{E}} = 0$ shows that the circular integer vortex solution is the stable solution.
In the case of $n = 2$, the circular integer vortex solution becomes unstable against the vortex molecule solution at $\theta \sim 85^\circ$.
In the case of $n = 3$, the circular integer vortex solution changed into the rod-shaped vortex molecule solution at $\theta \sim 80^\circ$.
While the only rod-shaped vortex molecule appears as the stable solution at $80^\circ \lesssim \theta \lesssim 85^\circ$, the triangular-shaped vortex molecule solution appears as the metastable solution at $\theta \gtrsim 85^\circ$.

\begin{figure}[htb]
\centering
\includegraphics[width=0.45\linewidth]{colorbox.pdf} \\
\begin{minipage}{0.24\linewidth}
{\footnotesize (a) $\bar{\mathcal{E}} = -2.719$} \\
\includegraphics[width=0.99\linewidth]{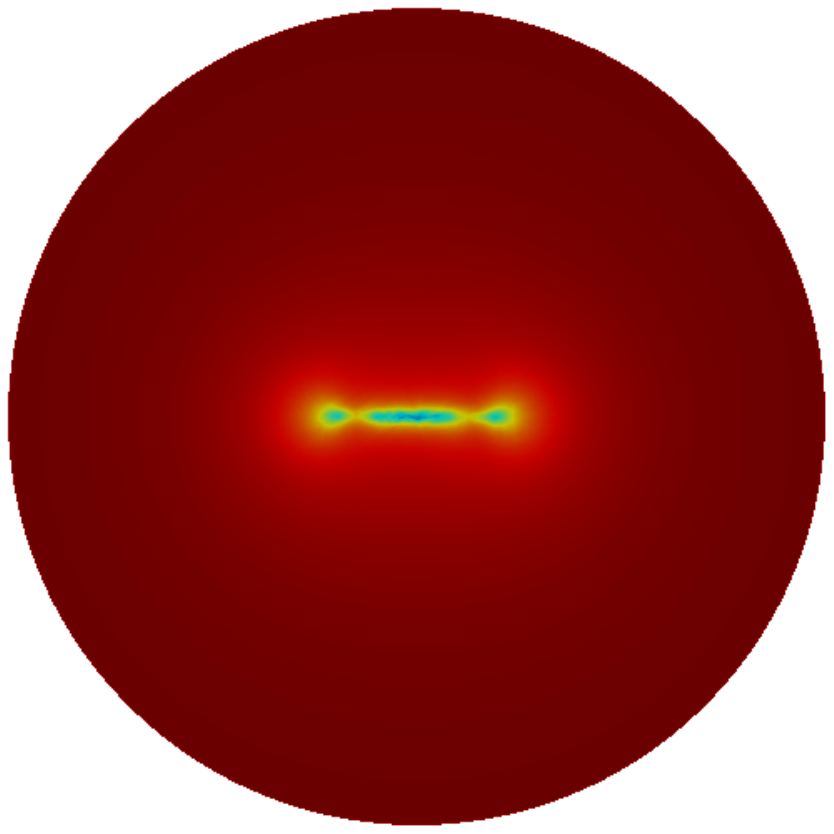}
\end{minipage}
\begin{minipage}{0.24\linewidth}
{\footnotesize (b) $\bar{\mathcal{E}} = -1.695$} \\
\includegraphics[width=0.99\linewidth]{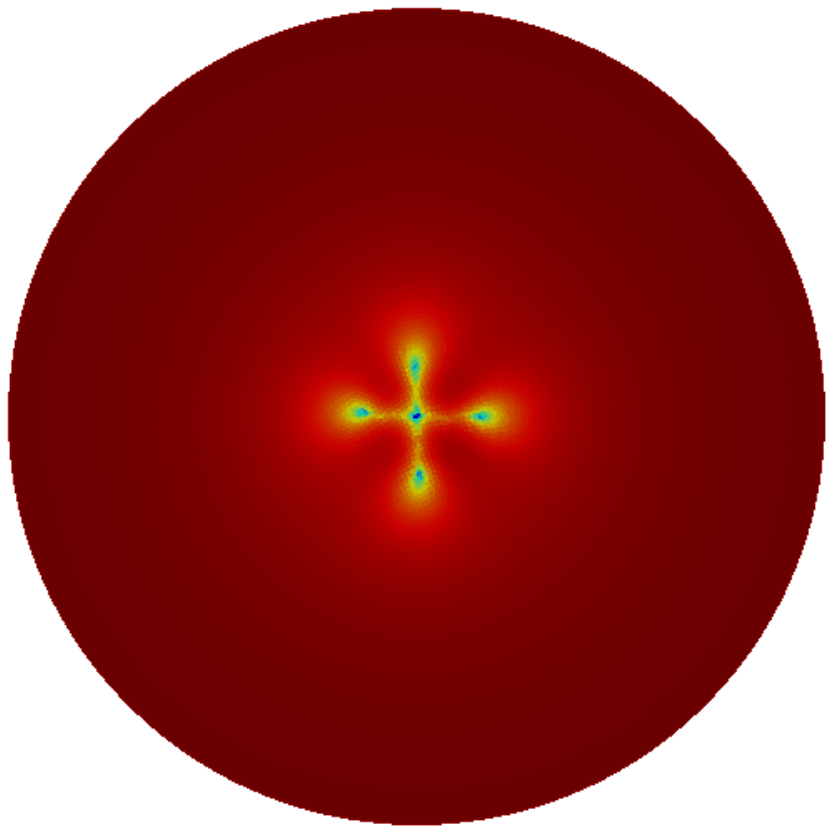}
\end{minipage}
\begin{minipage}{0.24\linewidth}
{\footnotesize (c) $\bar{\mathcal{E}} = -3.946$} \\
\includegraphics[width=0.99\linewidth]{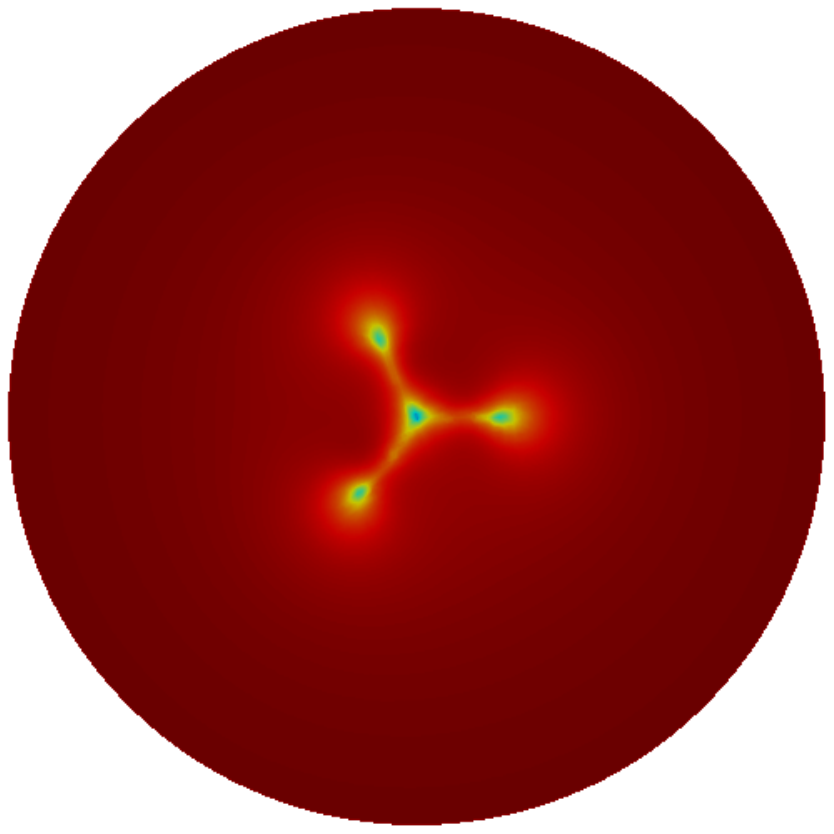}
\end{minipage}
\caption{
\label{fig:vortex-4}
Spatial configuration of $|\phi|^2$ for a integer vortex with $n = 4$.
The radius of the system is $40$.
}
\end{figure}
In the case of $n = 4$, there appear many metastable solutions and rod-shaped structure such as those shown in Figs. \ref{fig:vortex-2} (a) and \ref{fig:vortex-3} (a) is no more stable.
We show some of them in Fig. \ref{fig:vortex-4}.
The number of metastable solutions seems to be huge and 
it was impossible to exhaust all of them.
We will report the detailed analysis for $n \geq 4$ elsewhere.

%%%%%%%%%%%%%%%%%%%%%
\section{Summary and Discussion \label{sec:summary} }

%The modified XY model is a modification of the XY model 
%by addition of a half-periodic term.
As a generalization of the modified  XY  and Goldstone models,
 we have defined the ${\mathbb Z}_n$ modified XY and Goldstone models,
 having a $2 \pi/n$ periodic term in addition to the usual XY ($2 \pi$ periodic) term.
We have pointed out that the modified Goldstone model  
can be regarded as a nolinear sigma model with the target space 
of the  orbifold geometry 
${\mathbb C}/{\mathbb Z}_n$ 
with the orbifold singularity resolved.
Depending on the place of the vacua, we have found the two different schemes: the integer vortex sheme in which the XY term is dominant 
and the fractional vortex scheme in which the modified term is dominant.
%in which a vortex is an integer vortex.
We have exhausted vortex solutions for $n=2,3$ 
and have 
 fond a vortex confinement transition from an integer vortex 
 in the integer vortex scheme
 to a vortex molecle, i.e. $n$ $1/n$-quantized vortices connected by a domain wall (or walls) in the fractional vortex scheme.
In the case of $n=3$, we have found a Y shaped molecule as metastable solution and the most stable solutions is a rod-shaped molecule 
for the fractional vortex regime.

When two (or $n$) complex scalar fields coupled by a Josephson term(s)  
also admits a vortex molecule of half-quantized ($1/n$ quantized) vortices, 
see Refs. \cite{Son:2001td,Garcia:2002,Kasamatsu:2004tvg,Kasamatsu:2005,
Cipriani:2013nya,Nitta:2013eaa,Tylutki:2016mgy,Calderaro:2017,Eto:2017rfr} 
for two-components and Refs. 
\cite{Kuopanportti:2011,Eto:2012rc,Eto:2013spa,Cipriani:2013wia,Orlova:2016} for $n$ components.
This describes two-component BECs.
In this case, in contrast to the case of the modified model, there is no two-step phase transition 
\cite{Kobayashi:2018ezm,Kobayashi:2019}, 
but it is unclear what is a crucial difference between the two cases althouth the both admit similar solutions.

If we gauge the $U(1)$ symmetry of the Goldstone model, we have an Abelian-Higgs model. While the former admits a global vortex as we have discussed in this paper, the latter admits a local Abrikosov-Nielsen-Olesen vortex 
  \cite{Abrikosov:1956sx,Nielsen:1973cs}.
 It is an interesting question whether a modified Abelian-Higgs model admits local vortices of the molecule type. 
 There are several questions such as
whether there is a critical (BPS) coupling and whether it admits supersymmetric extension, see e.g. Ref.\cite{Eto:2006pg}.
 Whether there is any superconductor described by such models is also an interesting question.

\section*{Acknowledgements}

We would like to thank Chandrasekhar Chatterjee for a discussion at the early stage of this work, 
and him and Gergely Fej\H{o}s for collaboration of the previous paper. 
This work was supported by the Ministry of Education, Culture, Sports, Science (MEXT)-Supported Program for the Strategic Research Foundation at Private Universities ``Topological Science'' (Grant No. S1511006) 
and JSPS KAKENHI Grant Numbers 
16H03984 (M.\ K. and M.\ N.), 
This work is also supported in part by JSPS KAKENHI Grant Numbers 
%16H03984 (M.\ K. and M.\ N.), 
%19K14713 (C.\ C.), 
% and 
 18H01217 (M.\ N.).
%G.F. was also supported by the Hungarian National Research, Development and Innovation Office (Project No. 127982), and by the János Bolyai Research Scholarship of the Hungarian Academy of Sciences. 
The work of M.\ N. is also supported in part 
by a Grant-in-Aid for Scientific Research on Innovative Areas 
``Topological Materials Science" (KAKENHI Grant No. 15H05855) 
from MEXT of Japan.

%%%%%%%%%%%%%%%%%%%%%%%%%%%%%%%%%%%%%%%%%%%%%%%%%%%%%%%%%%%%
%\newpage

%%%%%%%%%% References %%%%%%%%%%%%%%%%%%%%%%%%%
\newcommand{\J}[4]{{\sl #1} {\bf #2} (#3) #4}
\newcommand{\andJ}[3]{{\bf #1} (#2) #3}
\newcommand{\AP}{Ann.\ Phys.\ (N.Y.)}
\newcommand{\MPL}{Mod.\ Phys.\ Lett.}
\newcommand{\NP}{Nucl.\ Phys.}
\newcommand{\PL}{Phys.\ Lett.}
\newcommand{\PR}{ Phys.\ Rev.}
\newcommand{\PRL}{Phys.\ Rev.\ Lett.}
\newcommand{\PTP}{Prog.\ Theor.\ Phys.}
\newcommand{\hep}[1]{{\tt hep-th/{#1}}}
%%%%%%%%%%%%%%%%%%%%%%%%%%%%%%%%%%%%%%%%%%%%%%%

\end{document}